\documentclass[aps,prl,twocolumn]{revtex4}

\usepackage{graphicx}
\usepackage{color}
\definecolor{red}{rgb}{0.3,0.3,0.3}
\usepackage{hyperref}
\usepackage[utf8]{inputenc}
\usepackage[T1]{fontenc}
\usepackage{scrextend}

\newcommand{\smin}{\scalebox{0.75}[1.0]{\( - \)}}

\begin{document}

\title{On the room temperature ferroelectricity of hydrogen-bonded charge transfer crystals}

\author{Gabriele D'Avino}
\author{Matthieu J. Verstraete}
\affiliation{Universit\'e de Li\`ege, Institut de Physique, Allée du 6 Ao\^ut 17, Sart-Tilman, B-4000 Li\`ege, Belgium and European Theoretical Spectroscopy Facility (\url{www.etsf.eu})}
\date{\today}

\begin{abstract}
We present a theoretical investigation of the anomalous ferroelectricity of mixed-stack charge transfer molecular crystals, based on the Peierls-Hubbard model, and first principles calculations for its  parameterization.
This approach is first validated by reproducing the temperature-induced transition and the electronic polarization of TTF-CA, and then applied to a novel series of hydrogen-bonded crystals, for which room temperature ferroelectricity has recently been claimed.
Our analysis shows that the hydrogen-bonded systems present a very low degree of charge transfer and hence support a very small  polarization.
A critical re-examination of experimental data supports our findings, shedding doubts on the ferroelectricity of these systems.
More generally, our modelling allows the rationalization of general features of the ferroelectric transition in charge transfer crystals, and suggests design principles for materials optimization.
\end{abstract}

\maketitle

Mixed-stack charge transfer (CT) crystals (e.g. TTF-CA, TTF-BA) are a spectacular example of multifunctionality in organic materials.
Being one of the few examples of quantum ferroelectricity among organics 
\cite{HoriuchiTokura_NatMat08,Horiuchi_Okimoto_Science03,Kobayashi_Horiuchi_PRL12,Kagawa_Horiuchi_PRL10}, CT crystals offer novel opportunities to achive magneto-electric control of the polarization \cite{KagawaHoriuchi_NatPhys10}, and for the realization of ultrafast nonlinear optical oscillators \cite{Miyamoto_NatComm13}.
Moreover, the occurrence of photoinduced phase transitions \cite{Collet_Science03}
triggered by multiexcitonic phenomena \cite{UemuraOkamoto_PRL10,MiyamotoKimura_PRL13} make these systems interesting for optical switching, memory and energy generation applications.

\begin{figure}[htbp]
\centering
{\includegraphics[trim=15 10 10 10,clip,width=0.45\textwidth]{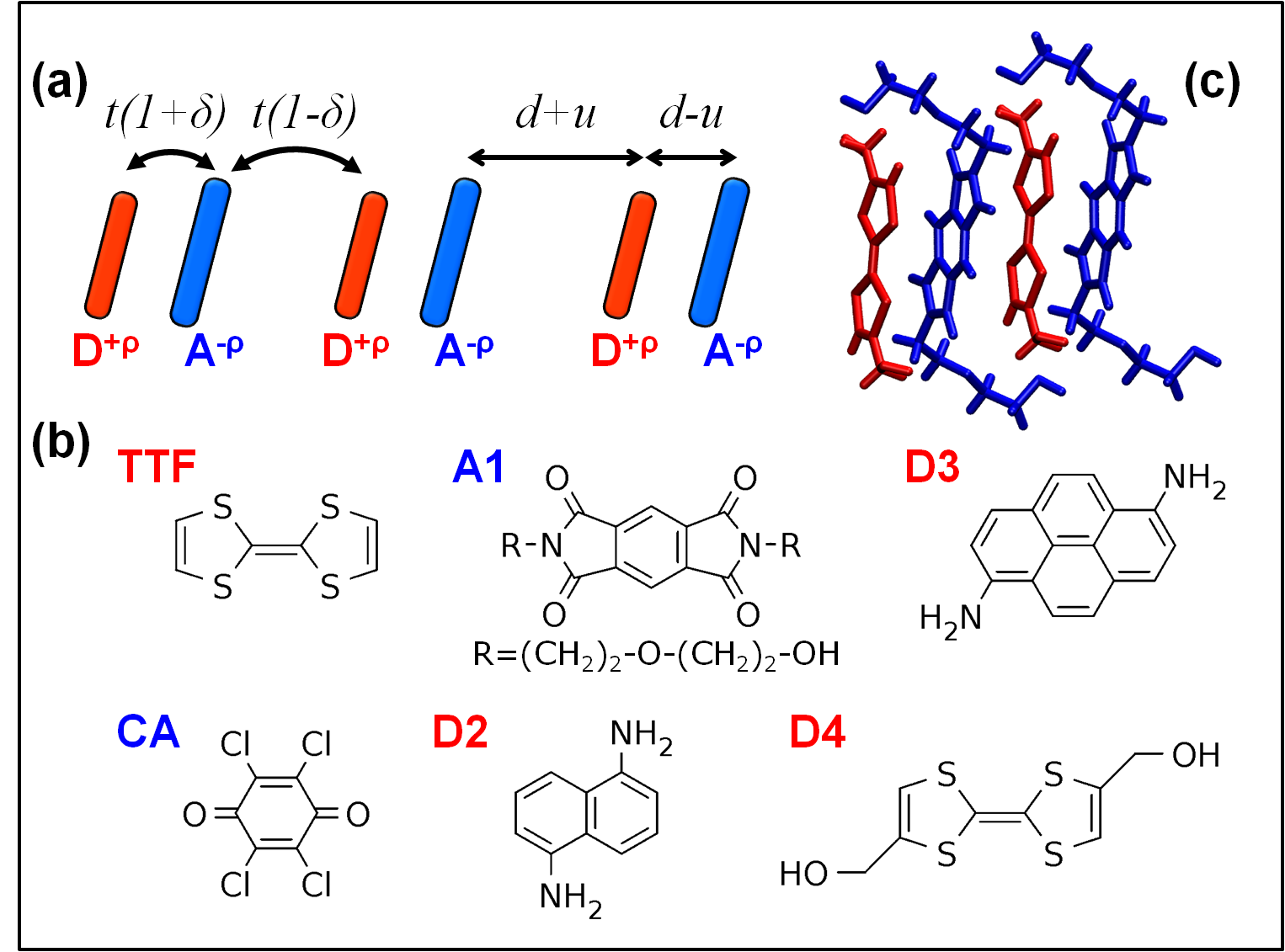} }
\caption{(color online) (a) Sketch of a dimerized mixed stack with alternating CT integrals and molecular displacements (exaggerated for clarity). 
(b) Chemical structures of D (red) and A (blue) molecules considered in this work. 
HBCT are complexes formed by the same acceptor, A1, and the three different donors, D1, D2 and D3.
(c) Perspective view of molecular packing in the A1-D4 crystal.}
\label{fig:fig1} 
\end{figure}

This intriguingly rich physics emerges from a quite simple structure, in which electron-donor (D) and -acceptor (A) molecules pack in an alternating one-dimensional (1D) pattern D$^{+\rho}$A$^{-\rho}$D$^{+\rho}$A$^{-\rho}$  characterized by a fractional charge transfer $\rho$ (see Figure \ref{fig:fig1}(a)).
Both neutral (N, $\rho\!\lesssim\!0.5$) and ionic (I, $\rho\!\gtrsim\!0.5$) CT crystals are known, and a few of them can undergo the so-called N-I transition, from a N phase to a low-temperature ($T$) and high-pressure I phase \cite{Torrance_PRL81_PNIT,Torrance_PRL81_TNIT,Girlando_Painelli_SynthMet04}.
In I systems a generalized Peierls instability may lead to a dimerization of the lattice, and ferroelectric phases characterized by an exceptionally strong electronic polarization, pointing antiparallel to molecular displacement dipoles \cite{Kobayashi_Horiuchi_PRL12}.

The archetypical organic CT ferroelectrics are the complexes of te\-tra\-tiafulvalene-halo-{\it p}-benzoquinone (TTF-QBr$_x$Cl$_{4-x}$) family, presenting transition temperatures $T_c=$81, 67 and 53 K for $\!x=\!0$ (TTF-CA), $x\!=\!1$ and $x\!=\!4$ (TTF-BA), respectively \cite{HoriuchiTokura_NatMat08}.
Room-temperature ferroelectricity has recently been reported in a novel series of CT crystals characterized by the presence of a supramolecular network of hydrogen-bonds (H-bonded charge transfer - HBCT, see Figure \ref{fig:fig1}) \cite{Tayi_Nature12}. 
This seems to pave the way for their application in realistic all-organic devices.
Remarkably, CT and H bonds, two phenomena possibly which both lead 
to ferroelectricity in molecular systems \cite{HoriuchiTokura_NatMat08,croco,benzi}, coexist in HBCT.

In this Letter, by means of a novel theoretical approach based on a model Hamiltonian fed with first principles inputs, we discuss on equal footing TTF-CA and HBCT, to determine the origin of the unprecedented properties of the latter and provide general insights on the anomalous ferroelectricity of mixed-stack CT crystals.
 
Electronic and structural instabilities of CT crystals are described by the 1D Modified Hubbard Hamiltonian with  electron-phonon coupling \cite{PainelliGirlando_PRB88,SoosPainelli_PRB07}, which, in conjunction with the modern theory of polarization in dielectrics \cite{Resta_PRL98}, provides a coherent framework that explains the divergence of the dielectric constant \cite{DelfreoPainelli_PRL02,SoosBewick_JChemPhys04}, vibrational spectra \cite{Girlando_Painelli_SynthMet04,DavinoMasino_PRB11} and diffuse X-ray data \cite{DavinoGirlando_PRL07}.
The physics of CT crystals is captured in its essence by the infinite-correlation limit ($U\!\rightarrow\!\infty$) of the Modified Peierls-Hubbard (MPH) Hamiltonian:
\begin{equation}\label{e:MH}
H = \Gamma_{\!e\!f\!f} \sum_i (\smin 1)^i\, \hat n_i \ - \  
t \sum_{i,\sigma} \left[ 1\! +\! (\smin 1)^i \delta \right] \hat b_i \ +\  \frac{N}{2\varepsilon_\delta} \delta^2 
\end{equation}
where $\Gamma_{\!e\!f\!f}$ is the effective ionization gap of a DA pair, $t$ is the CT integral, and $\delta$ is the dimensionless coordinate of the Peierls mode, with relaxation energy $\varepsilon_\delta$; 
$\hat b_i=(c^\dagger_{i,\sigma} c^{\phantom{\dagger}}_{i+1,\sigma} + H.c.)$ is the bond-order operator, and $N$ is the number of sites.
Hamiltonian (\ref{e:MH}) describes a continuous transition from a neutral band-insulating to an ionic Mott-insulating state upon decreasing $\Gamma_{\!e\!f\!f}$. 
At the critical point, characterized by a divergent polarizability, the instability to lattice  dimerization becomes unconditional, so that the I phase is always polar at $T=0$.

A more realistic model for CT crystals can be obtained by including electrostatic interactions in the 3D solid and intramolecular Holstein vibrations, 
allowing to describe discontinuous N-I transitions \cite{PainelliGirlando_PRB88,SoosPainelli_PRB07}.
A mean-field treatment of electrostatic interactions leads to the self-consistent 1D Hamiltonian: 
\begin{eqnarray}\label{e:Hct}
H_{CT} &=& (\Gamma + \frac{V}{2} +q +\varepsilon_c  \rho ) \sum_i (\smin 1)^i\, \hat n_i \nonumber \\
&- & t \sum_{i,\sigma} \left[ 1\! +\! (\smin 1)^i \delta \right] \hat b_i  
 \ +\  \frac{N}{2\varepsilon_\delta} \delta^2 \ + \ \frac{N}{2\varepsilon_q} q^2
\end{eqnarray}
plus constant terms \cite{SoosPainelli_PRB07}.
Here, $2\Gamma$ is the energy to ionize a DA pair at infinite distance and $V$ is the nearest neighbor electrostatic interaction; $q$ is the coordinate of the Holstein mode, with relaxation energy $\varepsilon_q$, and $\varepsilon_c\! =\! 2M\! -\! V$, where $M$ is the Madelung energy measuring the total strength of electrostatic interactions.

Traditionally, the model parameters were derived from experiments \cite{PainelliGirlando_JCP87,SoosPainelli_PRB07}, and empirical relationships between $\Gamma_{\!e\!f\!f}$ or $V$ and $T$ (or pressure) were established in order to induce the N-I transition \cite{DavinoGirlando_PRL07,DavinoMasino_PRB11}.
In this work we instead propose a novel and general approach to the model parameterization based on density functional theory (DFT). 
This strategy is first validated by reproducing the $T$-induced N-I transition of TTF-CA, 
and then applied to the series of HBCT crystals.

The values of $t$, $\delta$ and $\Gamma'=\Gamma +V/2$ are obtained by mapping DFT calculations on DA dimers to the corresponding effective model.
We perform DFT calculations on nearest neighbor DA dimers extracted from the crystal structures (at different $T$ for TTF-CA) \cite{Xrays_TTFCA,neutrons_TTFCA,Tayi_Nature12}, and compute energy and intermolecular CT in the singlet ($\rho_1$) and triplet ($\rho_3$) ground states.
We adopt three recent hybrid functionals, CAM-B3LYP, $\omega$B97X and M06-HF as implemented in the Gaussian09 suite \cite{g09}.
$\rho_1$ and $\rho_3$ were evaluated with natural population analysis atomic charges \cite{npa}. 

\begin{figure}[ht]
\centering
{\includegraphics[trim=25 25 25 25,clip,width=0.48\textwidth]{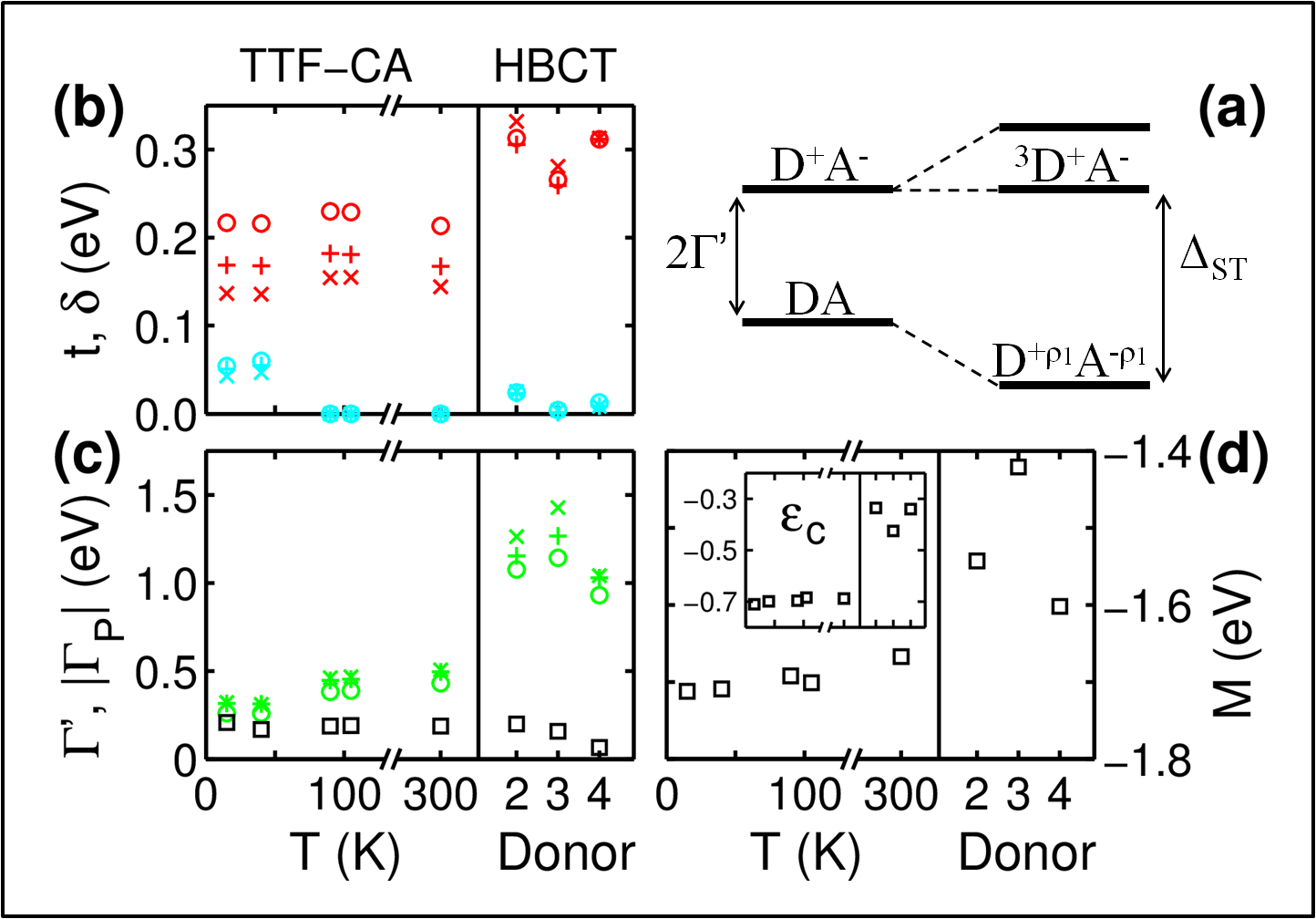}}
\caption{(color online) Model parameters calculated for TTF-CA and HBCT.
(a) Sketch of the energy levels for a DA dimer.
(b-c) Values of $t$ (red/dark gray symbols), $\delta$ (cyan/light gray) and $\Gamma'\!=\!\Gamma+V/2$ (green) obtained from CAM-B3LYP (circles), $\omega$B97X (pluses) and M06-HF (crosses) functionals and 6-31+G* basis set.
Squares correspond to the environment polarization contribution to the DA ionization energy, $\Gamma_P<0$.
(d) Madelung energy, $M$, and $\varepsilon_c$ (inset).}
\label{fig:MH_pars}
\end{figure}

On the other hand, in the strong-correlation limit (excluding double ionizations, D$^{2+}$A$^{2-}$, on physical grounds), the Modified Hubbard Hamiltonian for a DA dimer factorizes into a two-state model for the singlet subspace, plus three fully-CT ($\rho_3\!=\!1$) triplet states that are unaffected by the CT interaction (see Figure \ref{fig:MH_pars}(a)).
This simple analytical model is fully characterized by the two parameters $t$ and $\Gamma'$, whose values can be obtained from closed expressions in terms of the singlet-triplet gap, $\Delta_{ST}$, and the singlet ground-state CT, $\rho_1$, calculated with DFT.
By considering the symmetry inequivalent dimers in polar stacks one can access 
both $t$ and $\delta$ \cite{SM}.

The values of $t$, $\delta$ and $\Gamma'$ for TTF-CA and HBCT are shown in panels (b) and (c) of Figure \ref{fig:MH_pars}.
For TTF-CA we obtain a nearly $T$-independent  $t$ and comparable  results for different functionals, with CAM-B3LYP achieving a quantitative agreement with previous empirical estimates, $t\!\sim\!0.21$ eV \cite{PainelliGirlando_JCP87}. 
$\Gamma'$ shows a increasing trend with temperature, ascribable to the weakening of the nearest-neighbor interaction $V$ with the lattice expansion.

A very different scenario emerges for HBCT crystals, which are characterized by $t$ values comparable to TTF-CA, smaller dimerizations ($\delta\!<\!0.1\ t$) and, most importantly, values of $\Gamma'$ more than 0.5 eV larger than in TTF-CA.
The last result is clearly due to the weak DA character of HBCT complexes, as also confirmed by experimental \cite{Torrance_PRL81_PNIT,Tayi_Nature12} and calculated redox potentials \cite{SM}.

The other crucial parameters entering Hamiltonian (\ref{e:Hct}) are those quantifying electrostatic interactions in the solid. 
$M$ and $\varepsilon_c$ are here evaluated with a point-charge model, in which dielectric screening is accounted for with a microscopic model for molecular polarization, based on
DFT inputs  \cite{ts_prb01,davino_mecr,SM}.

The computed values of $M$, shown in Figure \ref{fig:MH_pars}(d), are large and negative as foreseeable for ionic lattices.
The Madelung energy decreases with $T$ in TTF-CA, confirming the expected 
gain in electrostatic energy upon lattice contraction. 
Smaller $M$ values are found for HBCT: 
the difference with respect to TTF-CA is due to the looser molecular packing in the presence of side chains.

The polarization of the environment is also responsible for a renormalization of the crystal ionization gap with respect to its gas-phase value, i.e. $\Gamma\rightarrow\Gamma +\Gamma_P$ \cite{SM}.
$\Gamma_P$ (squares in Figure \ref{fig:MH_pars}(c)) has been evaluated to be about -0.2 eV  \cite{SM}.

With the set of parameters at hand we can now perform MPH calculations specific for TTF-CA and HBCT. 
As in previous works \cite{SoosPainelli_PRB07,DavinoMasino_PRB11}, Hamiltonian (\ref{e:Hct}) is diagonalized exactly for chains with $N\!=\!16$ sites and periodic boundary conditions. 3D electrostatic interactions are treated at the mean-field level.
The Peierls phonon coordinate, $\delta$, is set to the values determined from experimental structures (see Figure \ref{fig:MH_pars}(b)), while the Holstein coordinate is relaxed  \cite{epsq}.
In the following, we will show results obtained with the CAM-B3LYP estimates of $t$, $\delta$ and $\Gamma$. The other functionals provide similar results \cite{SM}.  

\begin{figure}[t]
\centering
{\includegraphics[width=0.4\textwidth]{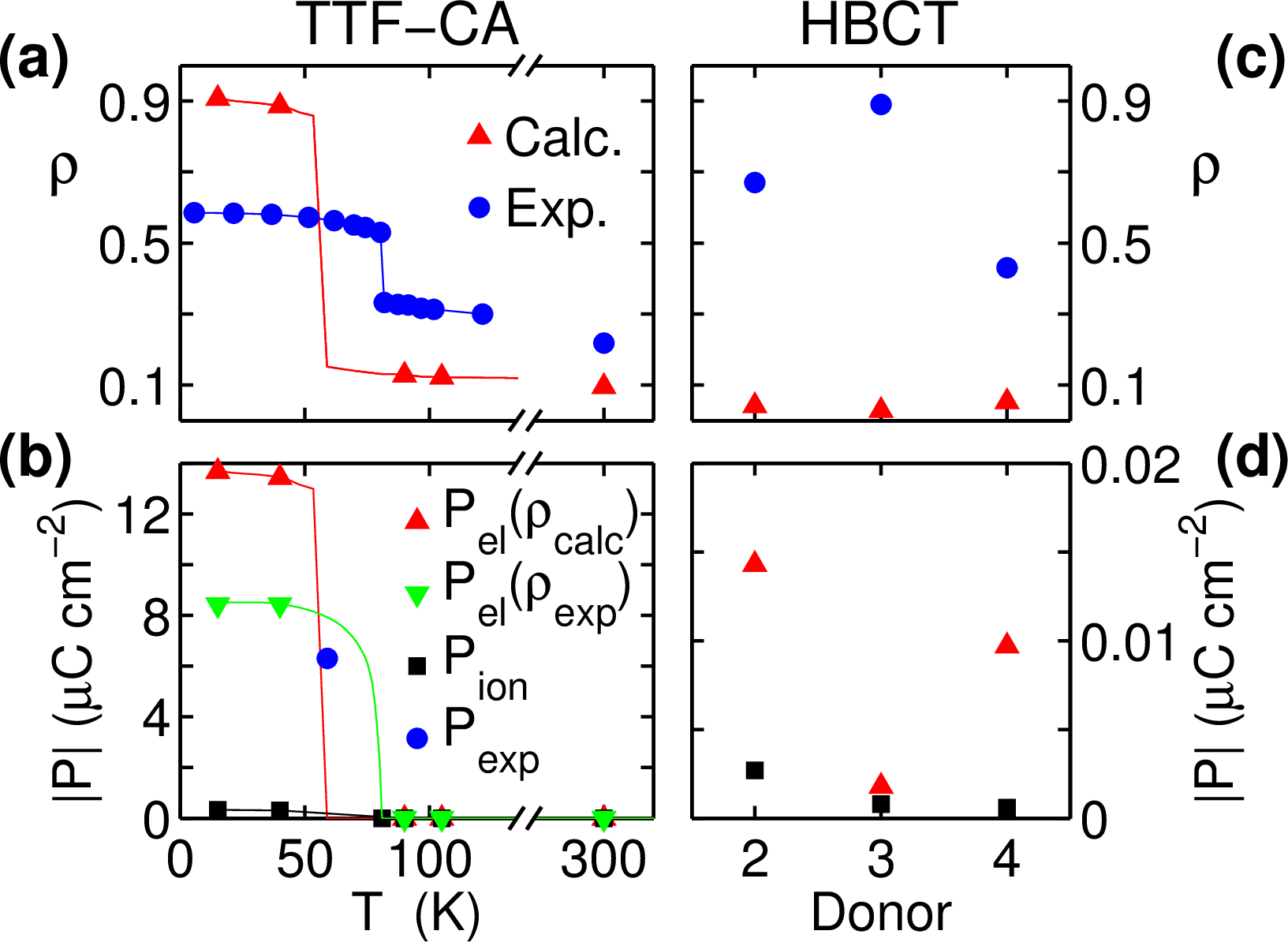}}
\caption{(color online) MPH results for ionicity and polarization of TTF-CA (at different $T$) and HBCT obtained with CAM-B3LYP parameters.
The calculations describe the $T$-induced N-I transition of TTF-CA (a), and the direction and magnitude of the polarization (b).
Unlike in Ref. \cite{Tayi_Nature12}, all HBCT crystals are found to be largely N (c) and with negligible $P$ (d) (see text).
Data points refer to calculations for parameters evaluated at experimental crystal structures. Higher $T$ resolution in TTF-CA is obtained by interpolating the model parameters in Figure \ref{fig:MH_pars}. }
\label{fig:rho_pol} 
\end{figure}

Calculated and experimental ionicity across the NIT of TTF-CA are shown in Figure \ref{fig:rho_pol}(a).
Our calculations correctly describe the first-order transition of TTF-CA and locate 
the critical point in the correct $T$ range.
Although the ionicity jump at the transition is overestimated, this result confirms the common picture of this transition: TTF-CA lies close to the N-I boundary, where a small increase in the Madelung energy drives the system from the N to the I phase.

The polarization along the stack (crystal axis $a$ in TTF-CA) can be decomposed in an electronic contribution, $P_{el}$, and an ionic one, $P_{ion}$.
According to the modern theory of polarization, $P_{el}$ is computed as a Berry phase \cite{Resta_PRL98,SoosBewick_JChemPhys04}:
\begin{equation}
P_{el}=\frac{ed}{\pi\Omega} \mathrm{Im} \ln \langle \Psi| \exp{\left(i\frac{2\pi \hat M}{N}\right)}| \Psi  \rangle 
\end{equation}
where $\Psi$ is the many-body ground state, $\hat M$ is the dipole moment operator of the open-boundary chain, $d$ is the intermolecular distance (at $\delta=0$), $\Omega$ is the volume per DA pair and $e$ the elementary charge.
The ionic contribution, due to frozen charges $\pm\rho$ at molecular sites displaced by $u$ (see Figure \ref{fig:fig1}(a)), is $P_{ion}\!=\!e\rho u/\Omega$.

The electronic polarization computed for TTF-CA, shown in Figure \ref{fig:rho_pol}(b),
is of the order of magnitude of experimental values ($6.3\ \mu\mathrm{C\, cm}^{-2}$ at 51 K \cite{Kobayashi_Horiuchi_PRL12}), and correctly points antiparallel to the almost negligible ionic contribution, as also reported by Picozzi and coworkers \cite{Giovannetti_Picozzi_PRL09}.
$P_{el}$ is evaluated at both the calculated and experimental ionicity, showing a very good agreement with experiments in the second case.
This allow us to conclude that, apart from inaccuracies in the estimation of $\rho$, the MPH model provides a quantitative description of the electronic polarization of CT crystals.
The better result obtained for $P_{el}$ with respect to the previous ab initio attempts \cite{Giovannetti_Picozzi_PRL09,Ishibashi_PhysB10}, suggests that an explicit, though approximate, treatment of the strong correlations seems to be more important than other details of the electronic structure.

MPH calculations for HBCT predict all the three crystals to be largely neutral ($\rho<0.1$) and characterized by very small polarizations (see Figure \ref{fig:rho_pol}(c) and (d)).
This is in marked contrast with the results of Ref. \cite{Tayi_Nature12}, where HBCT crystals were attributed $\rho$ values spanning a range of  0.4-0.9 and polarization comparable to or higher than TTF-CA.
The discrepancy between experiment and theory is addressed in the following.

Experimental estimates of $\rho$ in CT crystals rely on the linear dependence of the frequency of asymmetric C=O stretchings on the molecular charge, as it is well established for CA complexes 
\cite{Girlando_Marzola_JCP83}.
A similar procedure has been used for HBCT in Ref. \cite{Tayi_Nature12}, where a tiny \emph{hardening} of the C=O mode of A1, $\Delta\tilde\nu=\tilde\nu^{-}-\tilde\nu^{0}=14$ cm$^{-1}$, has been ascribed to the complete molecular ionization.
This is at odds with what is observed in CA, where, in agreement with chemical intuition, the relevant bond strongly \emph{weakens} in ionized molecules ($\Delta\tilde\nu=-160$ cm$^{-1}$)  \cite{Girlando_Marzola_JCP83}.
Normal mode calculations on neutral and charged molecules yield $\Delta\tilde\nu=\smin 198$ and $\smin 89$ cm$^{-1}$ in CA and A1, respectively \cite{SM}.
This leads us to conclude that the $\rho$ values of HBCT crystals are actually very similar and quite small.

Further confirmation of our ionicity estimates can be found by considering the whole experimental scenario. 
The relationship between $\rho$, the frequency of the CT optical transition, $\omega_{CT}$, and the difference between the redox potentials of D and A, $\Delta E_r$, has been established in Torrance's \emph{V-shaped} diagram, which has been empirically validated for many CT crystals \cite{Torrance_PRL81_PNIT}.
While TTF-CA lies  at the boundary between N and I phases, the experimental values reported for HBCT ($\omega_{CT}\sim 1.4-1.8$ eV and  $\Delta E_r\sim 1.1-1.6$ eV \cite{Tayi_Nature12}) safely locate these systems in the N region.

The polar crystal structures of HBCT were obtained by refining low-$T$ (84-100 K) X-ray diffraction data in non-centrosymmetric space groups. The only argument brought in support of the room-$T$ polarity of these systems is the violation of the mutual exclusion rule in infrared (IR) and Raman spectra \cite{Tayi_Nature12}.
However for HBCT there is no evidence of the very intense vibronic bands characterizing the IR spectra of dimerized stacks \cite{Girlando_Marzola_JCP83}. 
The coincidence of Raman and weak IR bands in busy vibrational spectra (because of the presence of side chains) should be considered accidental rather than a proof of the polarity of HBCT.

Finally, the polarization hysteresis loops of the HBCT crystals at room-$T$ 
show neither saturation nor reproducibility. 
These features are reminiscent of artifacts due to leakage current \cite{Scott_nature07,Scott_JPCM08}, which are indeed mentioned to occur in HBCT, especially at high $T$ \cite{Tayi_Nature12}.
Moreover, we note that the remnant polarization reported for A1-D4 is one order of magnitude higher that the upper limit prescribed by the modern theory of polarization for one-electron transfer.
Both theory and experiments cast doubts over the ferroelectricity of HBCT, calling for an unambiguous proof of the structure polarity and cleaner dielectric measurements.

\begin{figure}[htbp]
\centering
\includegraphics[width=0.47\textwidth]{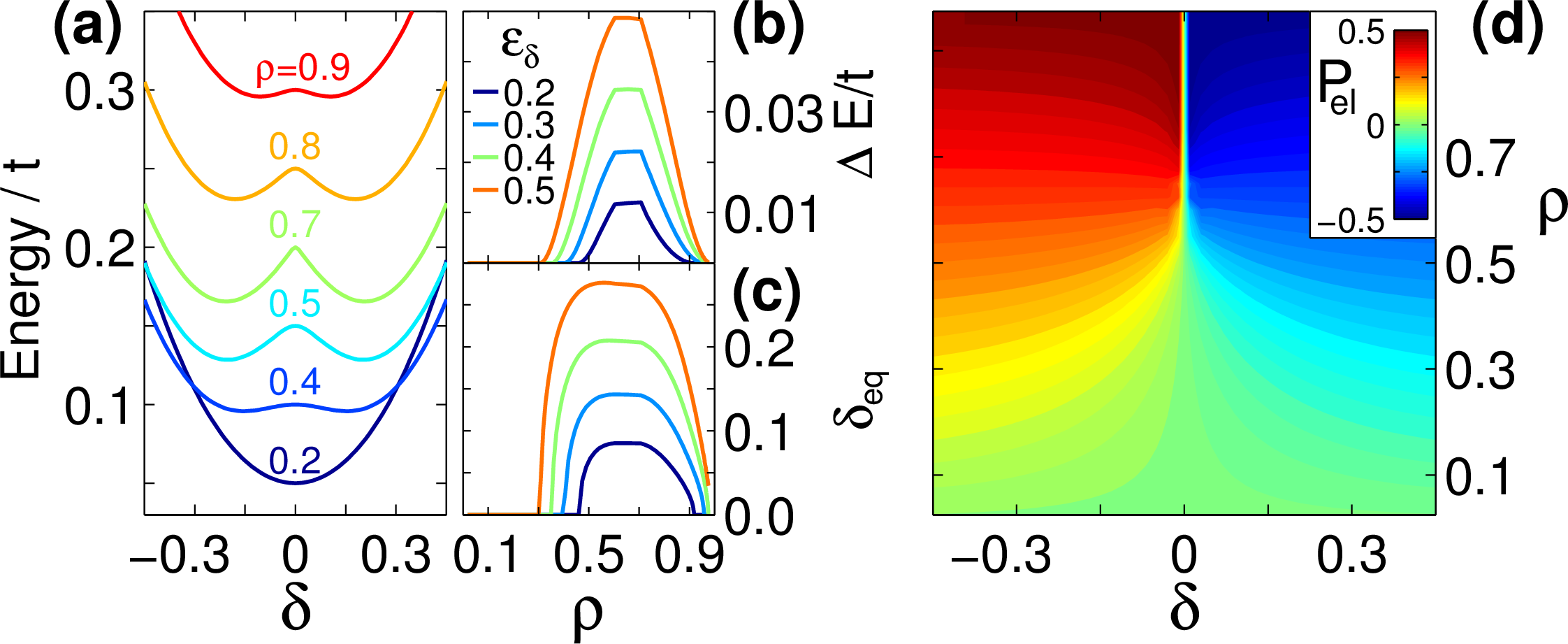} 
\caption{(a) Ground state energy of Hamiltonian (\ref{e:MH}) as a function of $\delta$ for different values of $\rho$  and $\varepsilon_\delta\!=\!0.4$.
(b) and (c) show the depth of the double well and the equilibrium position as a function of $\rho$ for different $\varepsilon_\delta$.
(d) Electronic polarization (dipole moment dipole per molecule in $ed$ units) of the MPH model as a function of $\rho$ and $\delta$. }
\label{fig:MH_FEtr} 
\end{figure}

In order to offer a comprehensive picture of the ferroelectric transition in CT crystals, we now present the general properties of the charge and lattice instability of the MPH model.
This provides useful guidelines for achieving robust ferroelectrics with high $T_c$.
As is well known, the increase of $\rho$ triggers the lattice instability and the ground-state potential in Figure \ref{fig:MH_FEtr}(a) develops a double well. The minima at $\pm\delta_{eq}$ correspond to polar phases of opposite polarization \cite{PainelliGirlando_PRB88,SoosPainelli_PRB07}.
The depth of the wells ($\Delta E$ in Figure \ref{fig:MH_FEtr}(b)) determines the stability of the polar phase against thermal fluctuations.
$\Delta E$ increases with the lattice softness and reaches a maximum at $\rho\!\sim\!0.6$.
In this regime, the lattice instability has a Peierls-like mechanism, with \emph{delocalized electrons} forming a bond-order charge density wave.
Conversely, the spin-Peierls instability of \emph{localized spins} in the Mott-insulating I phase, results in vanishing $\Delta E$ in the $\rho\!=\!1$ limit.
This explains why TTF-BA, featuring $\rho\!=\!0.95$  up to room $T$, dimerizes only below $53$ K, while higher transition temperatures are observed for TTF-CA and TTF-QBrCl$_3$, which both undergo ferroelectric transitions to I phases with $\rho\!\sim\!0.55$.

The magnitude of the electronic polarization is determined by the ionicity 
and the dimerization amplitude as shown in Figure \ref{fig:MH_FEtr}(d).
$P_{el}$ is an odd function of $\delta$ and vanishes by symmetry in the regular stack.
$P_{el}$  remains small at low ionicities, while, for $\rho\!>\!0.6$, it becomes large  
and discontinuous at $\delta\!=0$. 
Sizable CT ($\rho\!\gtrsim\!0.3$) is therefore an essential requisite to obtain a strong electronic polarization, which is instead less sensitive to the dimerization amplitude.
Large and nonlinear variations of $P_{el}$ with $\delta$ are signatures of the strong entanglement of correlated, yet delocalized, electrons with vibrations, suggesting that the concept of Born effective charges should be used with caution in these systems.

Hydrogen bonds can alter this picture, both by affecting the crystal packing (e.g. by favoring polar phases) or being responsible for an additional contribution to the polarization, as reported for several single- and multi-component H-bonded ferroelectrics \cite{HoriuchiTokura_NatMat08,croco,benzi}.
Our calculations account only for the structural effect on CT through the evaluation of the model parameters at experimental geometries.
The polarization of H-bonded ferroelectrics originates from the rearrangement of the $\pi$-conjugated electron system associated with the collective proton transfer, as also confirmed by DFT calculations  \cite{Ishii_Nagaosa_PRB06,croco}.
Since the weak H bonds of HBCT do not imply similar charge reorganization, we exclude that they could be responsible for the discrepancy between our estimate for the polarization and the data in Ref. \cite{Tayi_Nature12}.

In conclusion, we present a novel and general approach to the modelling of mixed-stack CT crystals, which is able to capture the anomalous electronic ferroelectricity of TTF-CA. 
We show that the novel HBCT complexes are all characterized by very low ionicities, 
as confirmed by a close reading of the original experimental data.
The latter do not unambiguously demonstrate the presence of ferroelectricity.
More generally, the theoretical framework we propose for CT crystals, by allowing one to target chemical specificity while fully accounting for the strong electronic correlations, is a powerful tool for the comprehension of the complex physics governing these promising multifunctional materials.

\begin{acknowledgements}
G.D. gratefully acknowledges enlightening discussions with Anna Painelli, Alberto Girlando and Matteo Masino.
We thank Eric Bousquet for carefully reading the manuscript.
This research was supported by the University of Liege and the EU in the context of the FP7-PEOPLE-COFUND-BEIPD project.
The authors acknowledge an A.R.C. grant (TheMoTherm 10/15-03) and a FRFC grant (Attosecond control of electron dynamics 2.4545.12) from the Communaut\'e Fran\c{c}aise de Belgique. Computer time was made available by PRACE-3IP project ThermoSpin on ARCHER (3IP FP7  RI-312763), Belgian CECI, and SEGI-ULg.
\end{acknowledgements}

\bibliographystyle{apsrev4-1}
\bibliography{hbct}

\begin{thebibliography}{37}%
\makeatletter
\providecommand \@ifxundefined [1]{%
 \@ifx{#1\undefined}
}%
\providecommand \@ifnum [1]{%
 \ifnum #1\expandafter \@firstoftwo
 \else \expandafter \@secondoftwo
 \fi
}%
\providecommand \@ifx [1]{%
 \ifx #1\expandafter \@firstoftwo
 \else \expandafter \@secondoftwo
 \fi
}%
\providecommand \natexlab [1]{#1}%
\providecommand \enquote  [1]{``#1''}%
\providecommand \bibnamefont  [1]{#1}%
\providecommand \bibfnamefont [1]{#1}%
\providecommand \citenamefont [1]{#1}%
\providecommand \href@noop [0]{\@secondoftwo}%
\providecommand \href [0]{\begingroup \@sanitize@url \@href}%
\providecommand \@href[1]{\@@startlink{#1}\@@href}%
\providecommand \@@href[1]{\endgroup#1\@@endlink}%
\providecommand \@sanitize@url [0]{\catcode `\\12\catcode `\$12\catcode
  `\&12\catcode `\#12\catcode `\^12\catcode `\_12\catcode `\%12\relax}%
\providecommand \@@startlink[1]{}%
\providecommand \@@endlink[0]{}%
\providecommand \url  [0]{\begingroup\@sanitize@url \@url }%
\providecommand \@url [1]{\endgroup\@href {#1}{\urlprefix }}%
\providecommand \urlprefix  [0]{URL }%
\providecommand \Eprint [0]{\href }%
\providecommand \doibase [0]{http://dx.doi.org/}%
\providecommand \selectlanguage [0]{\@gobble}%
\providecommand \bibinfo  [0]{\@secondoftwo}%
\providecommand \bibfield  [0]{\@secondoftwo}%
\providecommand \translation [1]{[#1]}%
\providecommand \BibitemOpen [0]{}%
\providecommand \bibitemStop [0]{}%
\providecommand \bibitemNoStop [0]{.\EOS\space}%
\providecommand \EOS [0]{\spacefactor3000\relax}%
\providecommand \BibitemShut  [1]{\csname bibitem#1\endcsname}%
\let\auto@bib@innerbib\@empty
\bibitem [{\citenamefont {Horiuchi}\ and\ \citenamefont
  {Tokura}(2008)}]{HoriuchiTokura_NatMat08}%
  \BibitemOpen
  \bibfield  {author} {\bibinfo {author} {\bibfnamefont {S.}~\bibnamefont
  {Horiuchi}}\ and\ \bibinfo {author} {\bibfnamefont {Y.}~\bibnamefont
  {Tokura}},\ }\href {\doibase 10.1038/nmat2137} {\bibfield  {journal}
  {\bibinfo  {journal} {Nature Mater.}\ }\textbf {\bibinfo {volume} {7}},\
  \bibinfo {pages} {357} (\bibinfo {year} {2008})}\BibitemShut {NoStop}%
\bibitem [{\citenamefont {Horiuchi}\ \emph {et~al.}(2003)\citenamefont
  {Horiuchi}, \citenamefont {Okimoto}, \citenamefont {Kumai},\ and\
  \citenamefont {Tokura}}]{Horiuchi_Okimoto_Science03}%
  \BibitemOpen
  \bibfield  {author} {\bibinfo {author} {\bibfnamefont {S.}~\bibnamefont
  {Horiuchi}}, \bibinfo {author} {\bibfnamefont {Y.}~\bibnamefont {Okimoto}},
  \bibinfo {author} {\bibfnamefont {R.}~\bibnamefont {Kumai}}, \ and\ \bibinfo
  {author} {\bibfnamefont {Y.}~\bibnamefont {Tokura}},\ }\href {\doibase
  10.1126/science.1076129} {\bibfield  {journal} {\bibinfo  {journal}
  {Science}\ }\textbf {\bibinfo {volume} {299}},\ \bibinfo {pages} {229}
  (\bibinfo {year} {2003})}\BibitemShut {NoStop}%
\bibitem [{\citenamefont {Kobayashi}\ \emph {et~al.}(2012)\citenamefont
  {Kobayashi}, \citenamefont {Horiuchi}, \citenamefont {Kumai}, \citenamefont
  {Kagawa}, \citenamefont {Murakami},\ and\ \citenamefont
  {Tokura}}]{Kobayashi_Horiuchi_PRL12}%
  \BibitemOpen
  \bibfield  {author} {\bibinfo {author} {\bibfnamefont {K.}~\bibnamefont
  {Kobayashi}}, \bibinfo {author} {\bibfnamefont {S.}~\bibnamefont {Horiuchi}},
  \bibinfo {author} {\bibfnamefont {R.}~\bibnamefont {Kumai}}, \bibinfo
  {author} {\bibfnamefont {F.}~\bibnamefont {Kagawa}}, \bibinfo {author}
  {\bibfnamefont {Y.}~\bibnamefont {Murakami}}, \ and\ \bibinfo {author}
  {\bibfnamefont {Y.}~\bibnamefont {Tokura}},\ }\href {\doibase
  10.1103/PhysRevLett.108.237601} {\bibfield  {journal} {\bibinfo  {journal}
  {Phys. Rev. Lett.}\ }\textbf {\bibinfo {volume} {108}},\ \bibinfo {pages}
  {237601} (\bibinfo {year} {2012})}\BibitemShut {NoStop}%
\bibitem [{\citenamefont {Kagawa}\ \emph
  {et~al.}(2010{\natexlab{a}})\citenamefont {Kagawa}, \citenamefont {Horiuchi},
  \citenamefont {Matsui}, \citenamefont {Kumai}, \citenamefont {Onose},
  \citenamefont {Hasegawa},\ and\ \citenamefont
  {Tokura}}]{Kagawa_Horiuchi_PRL10}%
  \BibitemOpen
  \bibfield  {author} {\bibinfo {author} {\bibfnamefont {F.}~\bibnamefont
  {Kagawa}}, \bibinfo {author} {\bibfnamefont {S.}~\bibnamefont {Horiuchi}},
  \bibinfo {author} {\bibfnamefont {H.}~\bibnamefont {Matsui}}, \bibinfo
  {author} {\bibfnamefont {R.}~\bibnamefont {Kumai}}, \bibinfo {author}
  {\bibfnamefont {Y.}~\bibnamefont {Onose}}, \bibinfo {author} {\bibfnamefont
  {T.}~\bibnamefont {Hasegawa}}, \ and\ \bibinfo {author} {\bibfnamefont
  {Y.}~\bibnamefont {Tokura}},\ }\href {\doibase
  10.1103/PhysRevLett.104.227602} {\bibfield  {journal} {\bibinfo  {journal}
  {Phys. Rev. Lett.}\ }\textbf {\bibinfo {volume} {104}},\ \bibinfo {pages}
  {227602} (\bibinfo {year} {2010}{\natexlab{a}})}\BibitemShut {NoStop}%
\bibitem [{\citenamefont {Kagawa}\ \emph
  {et~al.}(2010{\natexlab{b}})\citenamefont {Kagawa}, \citenamefont {Horiuchi},
  \citenamefont {Tokunaga}, \citenamefont {Fujioka},\ and\ \citenamefont
  {Tokura}}]{KagawaHoriuchi_NatPhys10}%
  \BibitemOpen
  \bibfield  {author} {\bibinfo {author} {\bibfnamefont {F.}~\bibnamefont
  {Kagawa}}, \bibinfo {author} {\bibfnamefont {S.}~\bibnamefont {Horiuchi}},
  \bibinfo {author} {\bibfnamefont {M.}~\bibnamefont {Tokunaga}}, \bibinfo
  {author} {\bibfnamefont {J.}~\bibnamefont {Fujioka}}, \ and\ \bibinfo
  {author} {\bibfnamefont {Y.}~\bibnamefont {Tokura}},\ }\href {\doibase
  10.1038/nphys1503} {\bibfield  {journal} {\bibinfo  {journal} {Nature Phys.}\
  }\textbf {\bibinfo {volume} {6}},\ \bibinfo {pages} {169} (\bibinfo {year}
  {2010}{\natexlab{b}})}\BibitemShut {NoStop}%
\bibitem [{\citenamefont {Miyamoto}\ \emph
  {et~al.}(2013{\natexlab{a}})\citenamefont {Miyamoto}, \citenamefont {Yada},
  \citenamefont {Yamakawa},\ and\ \citenamefont
  {Okamoto}}]{Miyamoto_NatComm13}%
  \BibitemOpen
  \bibfield  {author} {\bibinfo {author} {\bibfnamefont {T.}~\bibnamefont
  {Miyamoto}}, \bibinfo {author} {\bibfnamefont {H.}~\bibnamefont {Yada}},
  \bibinfo {author} {\bibfnamefont {H.}~\bibnamefont {Yamakawa}}, \ and\
  \bibinfo {author} {\bibfnamefont {H.}~\bibnamefont {Okamoto}},\ }\href@noop
  {} {\bibfield  {journal} {\bibinfo  {journal} {Nature Comm.}\ }\textbf
  {\bibinfo {volume} {4}},\ \bibinfo {pages} {2586} (\bibinfo {year}
  {2013}{\natexlab{a}})}\BibitemShut {NoStop}%
\bibitem [{\citenamefont {Collet}\ \emph {et~al.}(2003)\citenamefont {Collet},
  \citenamefont {Lemée-Cailleau}, \citenamefont {Buron-Le~Cointe},
  \citenamefont {Cailleau}, \citenamefont {Wulff}, \citenamefont {Luty},
  \citenamefont {Koshihara}, \citenamefont {Meyer}, \citenamefont {Toupet},
  \citenamefont {Rabiller},\ and\ \citenamefont {Techert}}]{Collet_Science03}%
  \BibitemOpen
  \bibfield  {author} {\bibinfo {author} {\bibfnamefont {E.}~\bibnamefont
  {Collet}}, \bibinfo {author} {\bibfnamefont {M.-H.}\ \bibnamefont
  {Lemée-Cailleau}}, \bibinfo {author} {\bibfnamefont {M.}~\bibnamefont
  {Buron-Le~Cointe}}, \bibinfo {author} {\bibfnamefont {H.}~\bibnamefont
  {Cailleau}}, \bibinfo {author} {\bibfnamefont {M.}~\bibnamefont {Wulff}},
  \bibinfo {author} {\bibfnamefont {T.}~\bibnamefont {Luty}}, \bibinfo {author}
  {\bibfnamefont {S.-Y.}\ \bibnamefont {Koshihara}}, \bibinfo {author}
  {\bibfnamefont {M.}~\bibnamefont {Meyer}}, \bibinfo {author} {\bibfnamefont
  {L.}~\bibnamefont {Toupet}}, \bibinfo {author} {\bibfnamefont
  {P.}~\bibnamefont {Rabiller}}, \ and\ \bibinfo {author} {\bibfnamefont
  {S.}~\bibnamefont {Techert}},\ }\href {\doibase 10.1126/science.1082001}
  {\bibfield  {journal} {\bibinfo  {journal} {Science}\ }\textbf {\bibinfo
  {volume} {300}},\ \bibinfo {pages} {612} (\bibinfo {year}
  {2003})}\BibitemShut {NoStop}%
\bibitem [{\citenamefont {Uemura}\ and\ \citenamefont
  {Okamoto}(2010)}]{UemuraOkamoto_PRL10}%
  \BibitemOpen
  \bibfield  {author} {\bibinfo {author} {\bibfnamefont {H.}~\bibnamefont
  {Uemura}}\ and\ \bibinfo {author} {\bibfnamefont {H.}~\bibnamefont
  {Okamoto}},\ }\href {\doibase 10.1103/PhysRevLett.105.258302} {\bibfield
  {journal} {\bibinfo  {journal} {Phys. Rev. Lett.}\ }\textbf {\bibinfo
  {volume} {105}},\ \bibinfo {pages} {258302} (\bibinfo {year}
  {2010})}\BibitemShut {NoStop}%
\bibitem [{\citenamefont {Miyamoto}\ \emph
  {et~al.}(2013{\natexlab{b}})\citenamefont {Miyamoto}, \citenamefont {Kimura},
  \citenamefont {Hamamoto}, \citenamefont {Uemura}, \citenamefont {Yada},
  \citenamefont {Matsuzaki}, \citenamefont {Horiuchi},\ and\ \citenamefont
  {Okamoto}}]{MiyamotoKimura_PRL13}%
  \BibitemOpen
  \bibfield  {author} {\bibinfo {author} {\bibfnamefont {T.}~\bibnamefont
  {Miyamoto}}, \bibinfo {author} {\bibfnamefont {K.}~\bibnamefont {Kimura}},
  \bibinfo {author} {\bibfnamefont {T.}~\bibnamefont {Hamamoto}}, \bibinfo
  {author} {\bibfnamefont {H.}~\bibnamefont {Uemura}}, \bibinfo {author}
  {\bibfnamefont {H.}~\bibnamefont {Yada}}, \bibinfo {author} {\bibfnamefont
  {H.}~\bibnamefont {Matsuzaki}}, \bibinfo {author} {\bibfnamefont
  {S.}~\bibnamefont {Horiuchi}}, \ and\ \bibinfo {author} {\bibfnamefont
  {H.}~\bibnamefont {Okamoto}},\ }\href {\doibase
  10.1103/PhysRevLett.111.187801} {\bibfield  {journal} {\bibinfo  {journal}
  {Phys. Rev. Lett.}\ }\textbf {\bibinfo {volume} {111}},\ \bibinfo {pages}
  {187801} (\bibinfo {year} {2013}{\natexlab{b}})}\BibitemShut {NoStop}%
\bibitem [{\citenamefont {Torrance}\ \emph
  {et~al.}(1981{\natexlab{a}})\citenamefont {Torrance}, \citenamefont
  {Vazquez}, \citenamefont {Mayerle},\ and\ \citenamefont
  {Lee}}]{Torrance_PRL81_PNIT}%
  \BibitemOpen
  \bibfield  {author} {\bibinfo {author} {\bibfnamefont {J.~B.}\ \bibnamefont
  {Torrance}}, \bibinfo {author} {\bibfnamefont {J.~E.}\ \bibnamefont
  {Vazquez}}, \bibinfo {author} {\bibfnamefont {J.~J.}\ \bibnamefont
  {Mayerle}}, \ and\ \bibinfo {author} {\bibfnamefont {V.~Y.}\ \bibnamefont
  {Lee}},\ }\href {\doibase 10.1103/PhysRevLett.46.253} {\bibfield  {journal}
  {\bibinfo  {journal} {Phys. Rev. Lett.}\ }\textbf {\bibinfo {volume} {46}},\
  \bibinfo {pages} {253} (\bibinfo {year} {1981}{\natexlab{a}})}\BibitemShut
  {NoStop}%
\bibitem [{\citenamefont {Torrance}\ \emph
  {et~al.}(1981{\natexlab{b}})\citenamefont {Torrance}, \citenamefont
  {Girlando}, \citenamefont {Mayerle}, \citenamefont {Crowley}, \citenamefont
  {Lee}, \citenamefont {Batail},\ and\ \citenamefont
  {LaPlaca}}]{Torrance_PRL81_TNIT}%
  \BibitemOpen
  \bibfield  {author} {\bibinfo {author} {\bibfnamefont {J.~B.}\ \bibnamefont
  {Torrance}}, \bibinfo {author} {\bibfnamefont {A.}~\bibnamefont {Girlando}},
  \bibinfo {author} {\bibfnamefont {J.~J.}\ \bibnamefont {Mayerle}}, \bibinfo
  {author} {\bibfnamefont {J.~I.}\ \bibnamefont {Crowley}}, \bibinfo {author}
  {\bibfnamefont {V.~Y.}\ \bibnamefont {Lee}}, \bibinfo {author} {\bibfnamefont
  {P.}~\bibnamefont {Batail}}, \ and\ \bibinfo {author} {\bibfnamefont {S.~J.}\
  \bibnamefont {LaPlaca}},\ }\href {\doibase 10.1103/PhysRevLett.47.1747}
  {\bibfield  {journal} {\bibinfo  {journal} {Phys. Rev. Lett.}\ }\textbf
  {\bibinfo {volume} {47}},\ \bibinfo {pages} {1747} (\bibinfo {year}
  {1981}{\natexlab{b}})}\BibitemShut {NoStop}%
\bibitem [{\citenamefont {Girlando}\ \emph {et~al.}(2004)\citenamefont
  {Girlando}, \citenamefont {Painelli}, \citenamefont {Bewick},\ and\
  \citenamefont {Soos}}]{Girlando_Painelli_SynthMet04}%
  \BibitemOpen
  \bibfield  {author} {\bibinfo {author} {\bibfnamefont {A.}~\bibnamefont
  {Girlando}}, \bibinfo {author} {\bibfnamefont {A.}~\bibnamefont {Painelli}},
  \bibinfo {author} {\bibfnamefont {S.}~\bibnamefont {Bewick}}, \ and\ \bibinfo
  {author} {\bibfnamefont {Z.}~\bibnamefont {Soos}},\ }\href {\doibase
  http://dx.doi.org/10.1016/j.synthmet.2003.11.004} {\bibfield  {journal}
  {\bibinfo  {journal} {Synthetic Metals}\ }\textbf {\bibinfo {volume} {141}},\
  \bibinfo {pages} {129 } (\bibinfo {year} {2004})}\BibitemShut {NoStop}%
\bibitem [{\citenamefont {Tayi}\ \emph {et~al.}(2012)\citenamefont {Tayi},
  \citenamefont {Shveyd}, \citenamefont {Sue}, \citenamefont {Szarko},
  \citenamefont {Rolczynski}, \citenamefont {Cao}, \citenamefont {Kennedy},
  \citenamefont {Sarjeant}, \citenamefont {Stern}, \citenamefont {Paxton},
  \citenamefont {Wu}, \citenamefont {Dey}, \citenamefont {Fahrenbach},
  \citenamefont {Guest}, \citenamefont {Mohseni}, \citenamefont {Chen},
  \citenamefont {Wang}, \citenamefont {Stoddart},\ and\ \citenamefont
  {Stupp}}]{Tayi_Nature12}%
  \BibitemOpen
  \bibfield  {author} {\bibinfo {author} {\bibfnamefont {A.~S.}\ \bibnamefont
  {Tayi}}, \bibinfo {author} {\bibfnamefont {A.~K.}\ \bibnamefont {Shveyd}},
  \bibinfo {author} {\bibfnamefont {A.~C.-H.}\ \bibnamefont {Sue}}, \bibinfo
  {author} {\bibfnamefont {J.~M.}\ \bibnamefont {Szarko}}, \bibinfo {author}
  {\bibfnamefont {B.~S.}\ \bibnamefont {Rolczynski}}, \bibinfo {author}
  {\bibfnamefont {D.}~\bibnamefont {Cao}}, \bibinfo {author} {\bibfnamefont
  {T.~J.}\ \bibnamefont {Kennedy}}, \bibinfo {author} {\bibfnamefont {A.~A.}\
  \bibnamefont {Sarjeant}}, \bibinfo {author} {\bibfnamefont {C.~L.}\
  \bibnamefont {Stern}}, \bibinfo {author} {\bibfnamefont {W.~F.}\ \bibnamefont
  {Paxton}}, \bibinfo {author} {\bibfnamefont {W.}~\bibnamefont {Wu}}, \bibinfo
  {author} {\bibfnamefont {S.~K.}\ \bibnamefont {Dey}}, \bibinfo {author}
  {\bibfnamefont {A.~C.}\ \bibnamefont {Fahrenbach}}, \bibinfo {author}
  {\bibfnamefont {J.~R.}\ \bibnamefont {Guest}}, \bibinfo {author}
  {\bibfnamefont {H.}~\bibnamefont {Mohseni}}, \bibinfo {author} {\bibfnamefont
  {L.~X.}\ \bibnamefont {Chen}}, \bibinfo {author} {\bibfnamefont {K.~L.}\
  \bibnamefont {Wang}}, \bibinfo {author} {\bibfnamefont {J.~F.}\ \bibnamefont
  {Stoddart}}, \ and\ \bibinfo {author} {\bibfnamefont {S.~I.}\ \bibnamefont
  {Stupp}},\ }\href {\doibase 10.1038/nature11395} {\bibfield  {journal}
  {\bibinfo  {journal} {Nature}\ }\textbf {\bibinfo {volume} {488}},\ \bibinfo
  {pages} {485} (\bibinfo {year} {2012})}\BibitemShut {NoStop}%
\bibitem [{\citenamefont {Horiuchi}\ \emph {et~al.}(2010)\citenamefont
  {Horiuchi}, \citenamefont {Tokunaga}, \citenamefont {Giovannetti},
  \citenamefont {Picozzi}, \citenamefont {Itoh}, \citenamefont {Shimano},
  \citenamefont {Kumai},\ and\ \citenamefont {Tokura}}]{croco}%
  \BibitemOpen
  \bibfield  {author} {\bibinfo {author} {\bibfnamefont {S.}~\bibnamefont
  {Horiuchi}}, \bibinfo {author} {\bibfnamefont {Y.}~\bibnamefont {Tokunaga}},
  \bibinfo {author} {\bibfnamefont {G.}~\bibnamefont {Giovannetti}}, \bibinfo
  {author} {\bibfnamefont {S.}~\bibnamefont {Picozzi}}, \bibinfo {author}
  {\bibfnamefont {H.}~\bibnamefont {Itoh}}, \bibinfo {author} {\bibfnamefont
  {R.}~\bibnamefont {Shimano}}, \bibinfo {author} {\bibfnamefont
  {R.}~\bibnamefont {Kumai}}, \ and\ \bibinfo {author} {\bibfnamefont
  {Y.}~\bibnamefont {Tokura}},\ }\href {\doibase 10.1038/nature0873} {\bibfield
   {journal} {\bibinfo  {journal} {Nature}\ }\textbf {\bibinfo {volume}
  {463}},\ \bibinfo {pages} {789} (\bibinfo {year} {2010})}\BibitemShut
  {NoStop}%
\bibitem [{\citenamefont {Horiuchi}\ \emph {et~al.}(2012)\citenamefont
  {Horiuchi}, \citenamefont {Kagawa}, \citenamefont {Hatahara}, \citenamefont
  {Kobayashi}, \citenamefont {Kumai}, \citenamefont {Murakami},\ and\
  \citenamefont {Tokura}}]{benzi}%
  \BibitemOpen
  \bibfield  {author} {\bibinfo {author} {\bibfnamefont {S.}~\bibnamefont
  {Horiuchi}}, \bibinfo {author} {\bibfnamefont {F.}~\bibnamefont {Kagawa}},
  \bibinfo {author} {\bibfnamefont {K.}~\bibnamefont {Hatahara}}, \bibinfo
  {author} {\bibfnamefont {K.}~\bibnamefont {Kobayashi}}, \bibinfo {author}
  {\bibfnamefont {R.}~\bibnamefont {Kumai}}, \bibinfo {author} {\bibfnamefont
  {Y.}~\bibnamefont {Murakami}}, \ and\ \bibinfo {author} {\bibfnamefont
  {Y.}~\bibnamefont {Tokura}},\ }\href {\doibase 10.1038/ncomms2322} {\bibfield
   {journal} {\bibinfo  {journal} {Nature Communications}\ }\textbf {\bibinfo
  {volume} {3}},\ \bibinfo {pages} {1308} (\bibinfo {year} {2012})}\BibitemShut
  {NoStop}%
\bibitem [{\citenamefont {Painelli}\ and\ \citenamefont
  {Girlando}(1988)}]{PainelliGirlando_PRB88}%
  \BibitemOpen
  \bibfield  {author} {\bibinfo {author} {\bibfnamefont {A.}~\bibnamefont
  {Painelli}}\ and\ \bibinfo {author} {\bibfnamefont {A.}~\bibnamefont
  {Girlando}},\ }\href {\doibase 10.1103/PhysRevB.37.5748} {\bibfield
  {journal} {\bibinfo  {journal} {Phys. Rev. B}\ }\textbf {\bibinfo {volume}
  {37}},\ \bibinfo {pages} {5748} (\bibinfo {year} {1988})}\BibitemShut
  {NoStop}%
\bibitem [{\citenamefont {Soos}\ and\ \citenamefont
  {Painelli}(2007)}]{SoosPainelli_PRB07}%
  \BibitemOpen
  \bibfield  {author} {\bibinfo {author} {\bibfnamefont {Z.~G.}\ \bibnamefont
  {Soos}}\ and\ \bibinfo {author} {\bibfnamefont {A.}~\bibnamefont
  {Painelli}},\ }\href {\doibase 10.1103/PhysRevB.75.155119} {\bibfield
  {journal} {\bibinfo  {journal} {Phys. Rev. B}\ }\textbf {\bibinfo {volume}
  {75}},\ \bibinfo {pages} {155119} (\bibinfo {year} {2007})}\BibitemShut
  {NoStop}%
\bibitem [{\citenamefont {Resta}(1998)}]{Resta_PRL98}%
  \BibitemOpen
  \bibfield  {author} {\bibinfo {author} {\bibfnamefont {R.}~\bibnamefont
  {Resta}},\ }\href {\doibase 10.1103/PhysRevLett.80.1800} {\bibfield
  {journal} {\bibinfo  {journal} {Phys. Rev. Lett.}\ }\textbf {\bibinfo
  {volume} {80}},\ \bibinfo {pages} {1800} (\bibinfo {year}
  {1998})}\BibitemShut {NoStop}%
\bibitem [{\citenamefont {Del~Freo}\ \emph {et~al.}(2002)\citenamefont
  {Del~Freo}, \citenamefont {Painelli},\ and\ \citenamefont
  {Soos}}]{DelfreoPainelli_PRL02}%
  \BibitemOpen
  \bibfield  {author} {\bibinfo {author} {\bibfnamefont {L.}~\bibnamefont
  {Del~Freo}}, \bibinfo {author} {\bibfnamefont {A.}~\bibnamefont {Painelli}},
  \ and\ \bibinfo {author} {\bibfnamefont {Z.~G.}\ \bibnamefont {Soos}},\
  }\href {\doibase 10.1103/PhysRevLett.89.027402} {\bibfield  {journal}
  {\bibinfo  {journal} {Phys. Rev. Lett.}\ }\textbf {\bibinfo {volume} {89}},\
  \bibinfo {pages} {027402} (\bibinfo {year} {2002})}\BibitemShut {NoStop}%
\bibitem [{\citenamefont {Soos}\ \emph {et~al.}(2004)\citenamefont {Soos},
  \citenamefont {Bewick}, \citenamefont {Peri},\ and\ \citenamefont
  {Painelli}}]{SoosBewick_JChemPhys04}%
  \BibitemOpen
  \bibfield  {author} {\bibinfo {author} {\bibfnamefont {Z.~G.}\ \bibnamefont
  {Soos}}, \bibinfo {author} {\bibfnamefont {S.~A.}\ \bibnamefont {Bewick}},
  \bibinfo {author} {\bibfnamefont {A.}~\bibnamefont {Peri}}, \ and\ \bibinfo
  {author} {\bibfnamefont {A.}~\bibnamefont {Painelli}},\ }\href {\doibase
  http://dx.doi.org/10.1063/1.1665824} {\bibfield  {journal} {\bibinfo
  {journal} {J. Chem. Phys.}\ }\textbf {\bibinfo {volume} {120}},\ \bibinfo
  {pages} {6712} (\bibinfo {year} {2004})}\BibitemShut {NoStop}%
\bibitem [{\citenamefont {D'Avino}\ \emph {et~al.}(2011)\citenamefont
  {D'Avino}, \citenamefont {Masino}, \citenamefont {Girlando},\ and\
  \citenamefont {Painelli}}]{DavinoMasino_PRB11}%
  \BibitemOpen
  \bibfield  {author} {\bibinfo {author} {\bibfnamefont {G.}~\bibnamefont
  {D'Avino}}, \bibinfo {author} {\bibfnamefont {M.}~\bibnamefont {Masino}},
  \bibinfo {author} {\bibfnamefont {A.}~\bibnamefont {Girlando}}, \ and\
  \bibinfo {author} {\bibfnamefont {A.}~\bibnamefont {Painelli}},\ }\href
  {\doibase 10.1103/PhysRevB.83.161105} {\bibfield  {journal} {\bibinfo
  {journal} {Phys. Rev. B}\ }\textbf {\bibinfo {volume} {83}},\ \bibinfo
  {pages} {161105} (\bibinfo {year} {2011})}\BibitemShut {NoStop}%
\bibitem [{\citenamefont {D'Avino}\ \emph {et~al.}(2007)\citenamefont
  {D'Avino}, \citenamefont {Girlando}, \citenamefont {Painelli}, \citenamefont
  {Lem\'ee-Cailleau},\ and\ \citenamefont {Soos}}]{DavinoGirlando_PRL07}%
  \BibitemOpen
  \bibfield  {author} {\bibinfo {author} {\bibfnamefont {G.}~\bibnamefont
  {D'Avino}}, \bibinfo {author} {\bibfnamefont {A.}~\bibnamefont {Girlando}},
  \bibinfo {author} {\bibfnamefont {A.}~\bibnamefont {Painelli}}, \bibinfo
  {author} {\bibfnamefont {M.-H.}\ \bibnamefont {Lem\'ee-Cailleau}}, \ and\
  \bibinfo {author} {\bibfnamefont {Z.~G.}\ \bibnamefont {Soos}},\ }\href
  {\doibase 10.1103/PhysRevLett.99.156407} {\bibfield  {journal} {\bibinfo
  {journal} {Phys. Rev. Lett.}\ }\textbf {\bibinfo {volume} {99}},\ \bibinfo
  {pages} {156407} (\bibinfo {year} {2007})}\BibitemShut {NoStop}%
\bibitem [{\citenamefont {Painelli}\ and\ \citenamefont
  {Girlando}(1987)}]{PainelliGirlando_JCP87}%
  \BibitemOpen
  \bibfield  {author} {\bibinfo {author} {\bibfnamefont {A.}~\bibnamefont
  {Painelli}}\ and\ \bibinfo {author} {\bibfnamefont {A.}~\bibnamefont
  {Girlando}},\ }\href {\doibase http://dx.doi.org/10.1063/1.453236} {\bibfield
   {journal} {\bibinfo  {journal} {J. Chem. Phys.}\ }\textbf {\bibinfo {volume}
  {87}},\ \bibinfo {pages} {1705} (\bibinfo {year} {1987})}\BibitemShut
  {NoStop}%
\bibitem [{\citenamefont {Garcia}\ \emph {et~al.}(2007)\citenamefont {Garcia},
  \citenamefont {Dahaoui}, \citenamefont {Katan}, \citenamefont {Souhassou},\
  and\ \citenamefont {Lecomte}}]{Xrays_TTFCA}%
  \BibitemOpen
  \bibfield  {author} {\bibinfo {author} {\bibfnamefont {P.}~\bibnamefont
  {Garcia}}, \bibinfo {author} {\bibfnamefont {S.}~\bibnamefont {Dahaoui}},
  \bibinfo {author} {\bibfnamefont {C.}~\bibnamefont {Katan}}, \bibinfo
  {author} {\bibfnamefont {M.}~\bibnamefont {Souhassou}}, \ and\ \bibinfo
  {author} {\bibfnamefont {C.}~\bibnamefont {Lecomte}},\ }\href {\doibase
  10.1039/B606642A} {\bibfield  {journal} {\bibinfo  {journal} {Faraday
  Discuss.}\ }\textbf {\bibinfo {volume} {135}},\ \bibinfo {pages} {217}
  (\bibinfo {year} {2007})}\BibitemShut {NoStop}%
\bibitem [{\citenamefont {Le~Cointe}\ \emph {et~al.}(1995)\citenamefont
  {Le~Cointe}, \citenamefont {Lem\'ee-Cailleau}, \citenamefont {Cailleau},
  \citenamefont {Toudic}, \citenamefont {Toupet}, \citenamefont {Heger},
  \citenamefont {Moussa}, \citenamefont {Schweiss}, \citenamefont {Kraft},\
  and\ \citenamefont {Karl}}]{neutrons_TTFCA}%
  \BibitemOpen
  \bibfield  {author} {\bibinfo {author} {\bibfnamefont {M.}~\bibnamefont
  {Le~Cointe}}, \bibinfo {author} {\bibfnamefont {M.~H.}\ \bibnamefont
  {Lem\'ee-Cailleau}}, \bibinfo {author} {\bibfnamefont {H.}~\bibnamefont
  {Cailleau}}, \bibinfo {author} {\bibfnamefont {B.}~\bibnamefont {Toudic}},
  \bibinfo {author} {\bibfnamefont {L.}~\bibnamefont {Toupet}}, \bibinfo
  {author} {\bibfnamefont {G.}~\bibnamefont {Heger}}, \bibinfo {author}
  {\bibfnamefont {F.}~\bibnamefont {Moussa}}, \bibinfo {author} {\bibfnamefont
  {P.}~\bibnamefont {Schweiss}}, \bibinfo {author} {\bibfnamefont {K.~H.}\
  \bibnamefont {Kraft}}, \ and\ \bibinfo {author} {\bibfnamefont
  {N.}~\bibnamefont {Karl}},\ }\href {\doibase 10.1103/PhysRevB.51.3374}
  {\bibfield  {journal} {\bibinfo  {journal} {Phys. Rev. B}\ }\textbf {\bibinfo
  {volume} {51}},\ \bibinfo {pages} {3374} (\bibinfo {year}
  {1995})}\BibitemShut {NoStop}%
\bibitem [{\citenamefont {Frisch}\ \emph {et~al.}()\citenamefont {Frisch} \emph
  {et~al.}}]{g09}%
  \BibitemOpen
  \bibfield  {author} {\bibinfo {author} {\bibfnamefont {M.~J.}\ \bibnamefont
  {Frisch}} \emph {et~al.},\ }\href@noop {} {\enquote {\bibinfo {title}
  {Gaussian09 {R}evision {D}.1},}\ }\bibinfo {note} {Gaussian Inc. Wallingford
  CT 2009}\BibitemShut {NoStop}%
\bibitem [{\citenamefont {Reed}\ \emph {et~al.}(1985)\citenamefont {Reed},
  \citenamefont {Weinstock},\ and\ \citenamefont {Weinhold}}]{npa}%
  \BibitemOpen
  \bibfield  {author} {\bibinfo {author} {\bibfnamefont {A.~E.}\ \bibnamefont
  {Reed}}, \bibinfo {author} {\bibfnamefont {R.~B.}\ \bibnamefont {Weinstock}},
  \ and\ \bibinfo {author} {\bibfnamefont {F.}~\bibnamefont {Weinhold}},\
  }\href {\doibase http://dx.doi.org/10.1063/1.449486} {\bibfield  {journal}
  {\bibinfo  {journal} {J. Chem. Phys.}\ }\textbf {\bibinfo {volume} {83}},\
  \bibinfo {pages} {735} (\bibinfo {year} {1985})}\BibitemShut {NoStop}%
\bibitem [{SM()}]{SM}%
  \BibitemOpen
  \href@noop {} {}\bibinfo {note} {See Supplemental Material for
  details.}\BibitemShut {Stop}%
\bibitem [{\citenamefont {Tsiper}\ and\ \citenamefont {Soos}(2001)}]{ts_prb01}%
  \BibitemOpen
  \bibfield  {author} {\bibinfo {author} {\bibfnamefont {E.~V.}\ \bibnamefont
  {Tsiper}}\ and\ \bibinfo {author} {\bibfnamefont {Z.~G.}\ \bibnamefont
  {Soos}},\ }\href {\doibase 10.1103/PhysRevB.64.195124} {\bibfield  {journal}
  {\bibinfo  {journal} {Phys. Rev. B}\ }\textbf {\bibinfo {volume} {64}},\
  \bibinfo {pages} {195124} (\bibinfo {year} {2001})}\BibitemShut {NoStop}%
\bibitem [{\citenamefont {D'Avino}\ \emph {et~al.}(2014)\citenamefont
  {D'Avino}, \citenamefont {Muccioli}, \citenamefont {Zannoni}, \citenamefont
  {Beljionne},\ and\ \citenamefont {Soos}}]{davino_mecr}%
  \BibitemOpen
  \bibfield  {author} {\bibinfo {author} {\bibfnamefont {G.}~\bibnamefont
  {D'Avino}}, \bibinfo {author} {\bibfnamefont {L.}~\bibnamefont {Muccioli}},
  \bibinfo {author} {\bibfnamefont {C.}~\bibnamefont {Zannoni}}, \bibinfo
  {author} {\bibfnamefont {D.}~\bibnamefont {Beljionne}}, \ and\ \bibinfo
  {author} {\bibfnamefont {Z.~G.}\ \bibnamefont {Soos}},\ }\href@noop {}
  {\enquote {\bibinfo {title} {Electronic polarization in organic crystals: a
  comparative study of induced dipoles and intramolecular charge redistribution
  schemes},}\ } (\bibinfo {year} {2014}),\ \bibinfo {note}
  {unpublished.}\BibitemShut {Stop}%
\bibitem [{eps()}]{epsq}%
  \BibitemOpen
  \href@noop {} {}\bibinfo {note} {$\varepsilon_q=0.40$, 0.56, 0.51 and 0.37 eV
  for TTF-CA, A1-D2, A1-D3 and A1-D4, respectively - B3LYP/6-31+G* estimates
  \cite{SM}.}\BibitemShut {Stop}%
\bibitem [{\citenamefont {Giovannetti}\ \emph {et~al.}(2009)\citenamefont
  {Giovannetti}, \citenamefont {Kumar}, \citenamefont {Stroppa}, \citenamefont
  {van~den Brink},\ and\ \citenamefont {Picozzi}}]{Giovannetti_Picozzi_PRL09}%
  \BibitemOpen
  \bibfield  {author} {\bibinfo {author} {\bibfnamefont {G.}~\bibnamefont
  {Giovannetti}}, \bibinfo {author} {\bibfnamefont {S.}~\bibnamefont {Kumar}},
  \bibinfo {author} {\bibfnamefont {A.}~\bibnamefont {Stroppa}}, \bibinfo
  {author} {\bibfnamefont {J.}~\bibnamefont {van~den Brink}}, \ and\ \bibinfo
  {author} {\bibfnamefont {S.}~\bibnamefont {Picozzi}},\ }\href {\doibase
  10.1103/PhysRevLett.103.266401} {\bibfield  {journal} {\bibinfo  {journal}
  {Phys. Rev. Lett.}\ }\textbf {\bibinfo {volume} {103}},\ \bibinfo {pages}
  {266401} (\bibinfo {year} {2009})}\BibitemShut {NoStop}%
\bibitem [{\citenamefont {Ishibashi}\ and\ \citenamefont
  {Terakura}(2010)}]{Ishibashi_PhysB10}%
  \BibitemOpen
  \bibfield  {author} {\bibinfo {author} {\bibfnamefont {S.}~\bibnamefont
  {Ishibashi}}\ and\ \bibinfo {author} {\bibfnamefont {K.}~\bibnamefont
  {Terakura}},\ }\href {\doibase http://dx.doi.org/10.1016/j.physb.2009.11.019}
  {\bibfield  {journal} {\bibinfo  {journal} {Physica B: Condensed Matter}\
  }\textbf {\bibinfo {volume} {405}},\ \bibinfo {pages} {S338 } (\bibinfo
  {year} {2010})}\BibitemShut {NoStop}%
\bibitem [{\citenamefont {Girlando}\ \emph {et~al.}(1983)\citenamefont
  {Girlando}, \citenamefont {Marzola}, \citenamefont {Pecile},\ and\
  \citenamefont {Torrance}}]{Girlando_Marzola_JCP83}%
  \BibitemOpen
  \bibfield  {author} {\bibinfo {author} {\bibfnamefont {A.}~\bibnamefont
  {Girlando}}, \bibinfo {author} {\bibfnamefont {F.}~\bibnamefont {Marzola}},
  \bibinfo {author} {\bibfnamefont {C.}~\bibnamefont {Pecile}}, \ and\ \bibinfo
  {author} {\bibfnamefont {J.~B.}\ \bibnamefont {Torrance}},\ }\href {\doibase
  http://dx.doi.org/10.1063/1.445833} {\bibfield  {journal} {\bibinfo
  {journal} {J. Chem. Phys.}\ }\textbf {\bibinfo {volume} {79}},\ \bibinfo
  {pages} {1075} (\bibinfo {year} {1983})}\BibitemShut {NoStop}%
\bibitem [{\citenamefont {Catalan}\ and\ \citenamefont
  {Scott}(2007)}]{Scott_nature07}%
  \BibitemOpen
  \bibfield  {author} {\bibinfo {author} {\bibfnamefont {G.}~\bibnamefont
  {Catalan}}\ and\ \bibinfo {author} {\bibfnamefont {J.~F.}\ \bibnamefont
  {Scott}},\ }\href {http://dx.doi.org/10.1038/nature06156} {\bibfield
  {journal} {\bibinfo  {journal} {Nature}\ }\textbf {\bibinfo {volume} {448}},\
  \bibinfo {pages} {E4} (\bibinfo {year} {2007})}\BibitemShut {NoStop}%
\bibitem [{\citenamefont {Scott}(2008)}]{Scott_JPCM08}%
  \BibitemOpen
  \bibfield  {author} {\bibinfo {author} {\bibfnamefont {J.~F.}\ \bibnamefont
  {Scott}},\ }\href {http://stacks.iop.org/0953-8984/20/i=2/a=021001}
  {\bibfield  {journal} {\bibinfo  {journal} {J. Phys. Condens. Mat.}\ }\textbf
  {\bibinfo {volume} {20}},\ \bibinfo {pages} {021001} (\bibinfo {year}
  {2008})}\BibitemShut {NoStop}%
\bibitem [{\citenamefont {Ishii}\ \emph {et~al.}(2006)\citenamefont {Ishii},
  \citenamefont {Nagaosa}, \citenamefont {Tokura},\ and\ \citenamefont
  {Terakura}}]{Ishii_Nagaosa_PRB06}%
  \BibitemOpen
  \bibfield  {author} {\bibinfo {author} {\bibfnamefont {F.}~\bibnamefont
  {Ishii}}, \bibinfo {author} {\bibfnamefont {N.}~\bibnamefont {Nagaosa}},
  \bibinfo {author} {\bibfnamefont {Y.}~\bibnamefont {Tokura}}, \ and\ \bibinfo
  {author} {\bibfnamefont {K.}~\bibnamefont {Terakura}},\ }\href {\doibase
  10.1103/PhysRevB.73.212105} {\bibfield  {journal} {\bibinfo  {journal} {Phys.
  Rev. B}\ }\textbf {\bibinfo {volume} {73}},\ \bibinfo {pages} {212105}
  (\bibinfo {year} {2006})}\BibitemShut {NoStop}%
\end{thebibliography}%

\end{document}